\documentclass[12pt]{extarticle} 
\usepackage[top=1in, bottom=1in,left=1in,right=1in]{geometry}

\usepackage[linesnumbered,ruled,lined]{algorithm2e}

\usepackage{makeidx,amsfonts, amsthm, amssymb, amsmath, graphicx, bbm, color, wrapfig, caption, subcaption, listings, float, mathtools, url, dsfont}
\usepackage{caption, subcaption}

\usepackage{hyperref}
\usepackage[capitalize,nameinlink]{cleveref}

\allowdisplaybreaks

\usepackage[shortlabels]{enumitem}
\setlist{leftmargin=*,itemsep=0.25\itemsep,parsep=0.35\parsep,topsep=0.25\topsep,partopsep=0.29\partopsep}

\newtheorem{theorem}{Theorem}
\newtheorem{other}{Theorem}
\newtheorem{definition}[other]{Definition}

\newtheorem{proposition}[other]{Proposition}
\newtheorem*{remark*}{Remarks}
\newtheorem{conjecture}[theorem]{Conjecture}

\renewenvironment{proof}[1][\proofname]{ { \it\bfseries #1: }}{\qed}

\newcommand{\cL}{{\mathcal{L}}}
\newcommand{\cN}{{\mathcal{N}}}

\newcommand{\field}[1]{\mathbb{#1}}
\newcommand{\R}{\field{R}}

\newcommand{\p}{\field{P}}
\newcommand{\1}{\mathbbm{1}}

\DeclareMathOperator{\E}{\mathbb{E}}

\newcommand{\mat}[2][rrrrrrrrrrrrrrrrrrrrrrrrrrrrrrrr]{\left[ \begin{array}{#1} #2 \\ \end{array}\right]}

\numberwithin{equation}{section}

\providecommand{\keywords}[1]
{
  \small	
  \textbf{\textit{Keywords: }} #1
}

\title{LoCoV: low dimension covariance voting  algorithm for portfolio optimization}
\author{Juntao Duan\footnote{Corresponding author. School of Mathematics, Georgia Institute of Technology, 
    \href{mailto:juntaoduan@gmail.com}{juntaoduan@gmail.com}}
    ,\quad  
    Ionel Popescu\footnote{ University of Bucharest, Faculty of Mathematics and Computer Science, Institute of Mathematics of the Romanian Academy, \href{mailto:ioionel@gmail.com }{ioionel@gmail.com}}
    }
\date{}
\begin{document}

\maketitle

\begin{abstract}
    Minimum-variance portfolio optimizations rely on accurate covariance estimator to obtain optimal portfolios. However, it usually suffers from large error from sample covariance matrix when the sample size $n$ is not significantly larger than the number of assets $p$. We analyze the random matrix aspects of portfolio optimization and identify the order of errors in sample optimal portfolio weight and show portfolio risk are underestimated when using samples. We also provide LoCoV (low dimension covariance voting) algorithm to reduce error inherited from random samples. From various experiments,  LoCoV is shown to outperform the classical method by a large margin.
\end{abstract}
\keywords{portfolio optimization; covariance; LoCoV; low dimension covariance voting }

\section{Introduction}
Portfolio theory pioneered by Markowitz in 1950's \cite{markowitz1952portfolio} is at the center of  theoretical developments in finance. The mean-variance model tells investors should hold a portfolio on the efficient frontier which trade off portfolio mean  (return) against variance (risk).  In practice, mean and variance are calculated using estimated sample mean and sample covariance matrix. However, estimation error in sample mean and covariance will significantly affect  the accuracy of the portfolio thus perform poorly in practice (see \cite{jobson1981putting, michaud1989markowitz}). Quantitative result on how sample covariance affects the performance are very limited. The bias in sample portfolio weight is discussed in \cite{el2010high} but no practical guidance is given on how large is the bias when use mean-variance model with sample data. We in this work will obtain that the order of magnitude of the error in sample  portfolio weight which is large when the sample size $n$ is comparable to the number of assets $p$. And the error decays in the rate of $\sqrt{\frac{p}{n}}$ as $n$ increases.

For this reason, there has been many work suggest different approaches to overcome standard mean-variance portfolio optimizations.  These suggestions include  imposing portfolio constraints (see \cite{jagannathan2003risk, demiguel2009generalized, behr2013portfolio}), use of factor models (\cite{chan1999portfolio}),  modifying objective to be more robust  (\cite{demiguel2009portfolio}) and improving sample covariance matrix estimation (\cite{ledoit2003improved}). Instead, in this work we use the observation from random matrix theory to provide alternative view on the error in sample covariance matrix. We propose LoCoV, low dimension covariance voting, which effectively exploits the accurate low dimensional covariance to vote on different assets. It outperform the standard sample portfolio by a large margin.

We shall first set up the problem. For simplicity, we only discuss minimum-variance portfolio optimization. Assume the true covariance admits diagonalization 
\[
\Sigma = P^T D^2 P
\]
where $D$ is a non-negative definite diagonal matrix, and $P$ is an orthogonal matrix. Then a data matrix (asset return) realized by random matrix $\cN (n\times p)$ with i.i.d. standard random variables is 
\[
X = \cN D P
\]
a sample covariance matrix is then obtained as 
\[
\hat{\Sigma} = P^TD\frac{\cN^T \cN}{n} DP
\]
We define the minimum variance portfolio to be the optimizer of 
\begin{equation}\label{eqn:true portfolio optimization}
    \begin{split}
		\min_w \;& w^T \Sigma \; w \\
		s.t. \;&  w^T \mathbbm{1} =1
    \end{split} 
\end{equation}
where $\mathbbm{1}= \mat[ccc]{1 & \cdots & 1}^T$. In reality, $\Sigma$ is not known, therefore it is replaced by an estimator $\hat{\Sigma}$ to obtain an approximated optimal portfolio. That is we solve
\begin{equation}\label{eqn:sample portfolio optimization}
\begin{split}
		\min_w \;& w^T \hat{\Sigma}\; w \\
		s.t. \;&  w^T \mathbbm{1} =1
\end{split} 
\end{equation}

\section{Universality of optimal portfolio weight and risk}
We first derive the solution of minimum-variance by the method of Lagrange multiplier since a closed form is available. Later on based on the explicit form of the solutions, we will investigate probabilistic properties of portfolio weight and risk.

Observe that  both $\Sigma$ and $\hat{\Sigma}$ take the form $A^T A$ where $A$ is $DP$ for true covariance $\Sigma$ and $A$ is $\frac{1}{\sqrt{n}}\cN DP$ for the sample covariance matrix $\hat{\Sigma}$. We shall define the portfolio optimization in the general form 
\begin{equation}\label{eqn:general form portfolio optimization}
    \begin{split}
		\min_w \;& w^T A^T A \; w \\
		s.t. \;&  w^T \mathbbm{1} =1
    \end{split} 
\end{equation}

Define the Lagrangian function
\[
\cL(w) = w^T A^TA w - \lambda (w^T \1 -1)
\]

Taking derivatives with respect to the portfolio weight $w$, and set  the gradient to be zero, 
\[
\nabla \cL = 2 w^TA^TA - \lambda \1^T =0
\]
write gradient as column vector this is 
\[
2A^TA w = \lambda \1
\]
For real life portfolio optimization, we can assume $A^TA$ ($\Sigma$ or $\hat{\Sigma}$) is invertible since otherwise optimal portfolio weight will have large error or ambiguity.  Then we  find the optimal portfolio weight 
\[
w = \frac{\lambda}{2} (A^TA)^{-1}\1
\]
We know the portfolio weights should be normalized so that they sum up to 1. Therefore $\lambda/2$ is essentially a normalizing factor. 
For convenience of notation, we make the following definition. 
\begin{definition}
    The \textbf{free (non-normalized) optimal weight} of portfolio optimization \ref{eqn:general form portfolio optimization}  is 
    \[
    S=  \mat[ccc]{s_1 & \cdots & s_p}^T = (A^TA)^{-1}\1  
    \]
    And denote its sum as 
    \[
    \|S\|_s := \sum_{k=1}^p s_k
    \]
\end{definition}
Normalizing the vector $S$ we obtain optimal portfolio weight 
\[
w^* = \frac{1}{\sum_{i=1}^p s_i} S = \frac{S}{\|S\|_s}
\]
It is easy to see $\lambda^* = \frac{2}{\sum_{i=1}^p s_i} = 2 \|S\|_s^{-1}$.
Then take dot product of $\nabla \cL$ and $w$, we  find 
\[
0= \nabla \cL^T w = 2 w^TA^TA w - \lambda \1^T w 
\]
and recall $\1^T w =1$, therefore, we find the minimum portfolio risk
\[
R(w^*) = w^{*T}A^TA w^* = \lambda^* /2 = \|S\|_s^{-1}
\]
We summarize the result as follows,

\begin{proposition}\label{prop:optimal formulas}
For the constrained optimization \ref{eqn:general form portfolio optimization}, the \textbf{free optimal weight}  is 
\begin{equation}\label{eqn:free optimal weight}
   S= \mat[c]{s_1\\\vdots \\ s_p} = (A^TA)^{-1}\1 
\end{equation}

Normalizing $S$, we obtain the \textbf{optimal portfolio weight} 
\begin{equation}\label{eqn:optimal weight}
    w^* = \|S\|_s^{-1} S
\end{equation}
and the \textbf{minimum portfolio risk} is 
\begin{equation}\label{eqn:optimal risk}
    R(w^*) =  \|S\|_s^{-1}
\end{equation}
where $\|S\|_s^{-1} = \sum_{k=1}^p s_k$.
\end{proposition}

\subsection{Behavior of sample portfolio}
Assume the diagonalization of true covariance matrix
\[
\Sigma =P^T D^2P =P^T diag (\sigma_1^2, \cdots, \sigma_p^2)P
\]
By proposition \ref{prop:optimal formulas}, plugging in $A^TA = \Sigma$, we find the \textbf{true free optimal weight} and \textbf{true optimal portfolio weight} of \ref{eqn:true portfolio optimization} are
\begin{equation}\label{eqn:true weight formula}
    S_{\Sigma} = \Sigma^{-1}\1 = P^T D^{-2}P \1 , \qquad w^* = \|S_{\Sigma} \|_s^{-1} S_{\Sigma}
\end{equation}

Then recall the return (data matrix) is generated as $X= \cN DP$ where $\cN$ is a $n\times p$ matrix with i.i.d. standard  random variables (mean zero and variance one). This leads to the sample covariance matrix 
\[
\hat{\Sigma} = P^T D\frac{\cN^T\cN}{n} D P
\]
Plugging in $A^TA = \hat{\Sigma}$ for proposition \ref{prop:optimal formulas}, we obtain \textbf{sample free optimal weight} and \textbf{sample optimal portfolio weight}
of \ref{eqn:sample portfolio optimization}
\begin{equation}\label{eqn:sample weight formula}
S_{\hat{\Sigma}} = P^T \hat{\Sigma}^{-1} P\1 = P^T D^{-1} \left( \frac{\cN^T\cN}{n}\right)^{-1} D^{-1}P \1 , \qquad \hat{w}^* = \|S_{\hat{\Sigma}}\|_s^{-1} S_{\hat{\Sigma}}
\end{equation}
The difference between $S_{\hat{\Sigma}}$ and $S_{\Sigma}$ depends on the random matrix (inverse of sample covariance) $M:=\left( \frac{\cN^T\cN}{n}\right)^{-1}$, diagonal matrix $D$ and orthogonal matrix $P$.   $M$ is the inverse of a sample covariance matrix. It is possible to directly use the formula for inverse from Cramer rule to analyze this random matrix and show $\E M = I$. Since this work mainly focus on improving the accuracy of portfolio, we will not pursue the probabilistic properties here (which shall be discussed in another work elsewhere).  Instead we use several experiments to show the sample portfolio weight $\hat{w}^*$ is centered around the true portfolio weight $w^*$.

\subsection{First example: sample portfolio of independent assets}

We shall start with the simplest case that all assets are independent, i.e. the matrix $P$ is identity. This means the true covariance matrix is a diagonal matrix 
$
\Sigma =D^2
$.
 Then by \ref{eqn:true weight formula} \textbf{true free optimal weight} and \textbf{true optimal portfolio weight}
\[
S_{\Sigma} = \Sigma^{-1}\1 = D^{-1}D^{-1}\1 = \mat[ccc]{\sigma_1^{-2} & \cdots & \sigma_p^{-2}}^T, \qquad w^* = \|S_{\Sigma} \|_s^{-1} S_{\Sigma} 
\]

Similarly by \ref{eqn:sample weight formula} \textbf{sample free optimal weight} and \textbf{sample optimal portfolio weight}
\[
S_{\hat{\Sigma}} = \hat{\Sigma}^{-1}\1 = D^{-1} \left( \frac{\cN^T\cN}{n}\right)^{-1} D^{-1}\1 , \qquad \hat{w}^* = \|S_{\hat{\Sigma}}\|_s^{-1} S_{\hat{\Sigma}}
\]

\begin{figure}[H]
     \centering
     \begin{subfigure}[b]{\textwidth}
         \centering
         \includegraphics[width=1\linewidth]{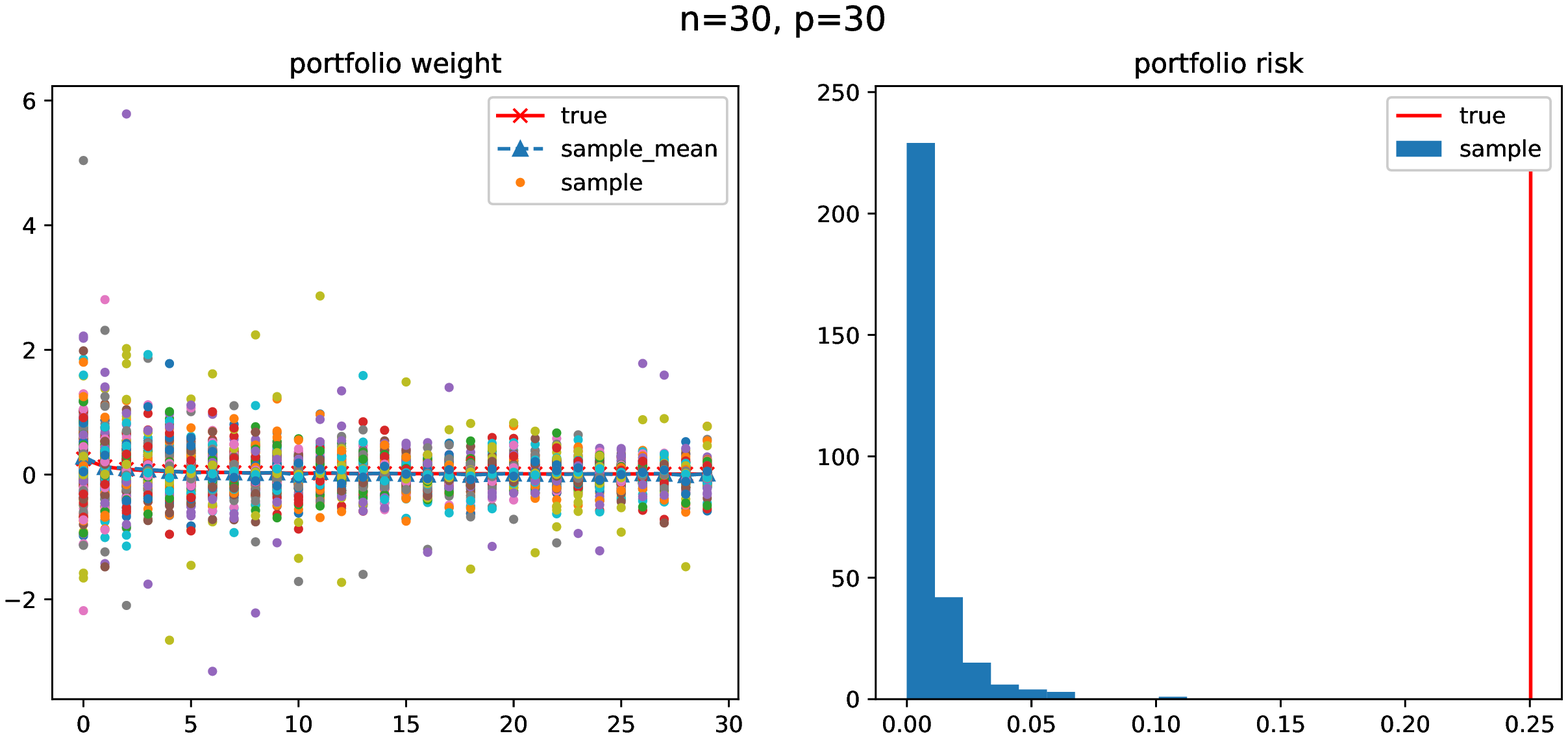}
     \end{subfigure} 
\end{figure}%
\begin{figure}[ht]\ContinuedFloat
     \begin{subfigure}[ht]{\textwidth}
         \centering
    \includegraphics[width=1\linewidth]{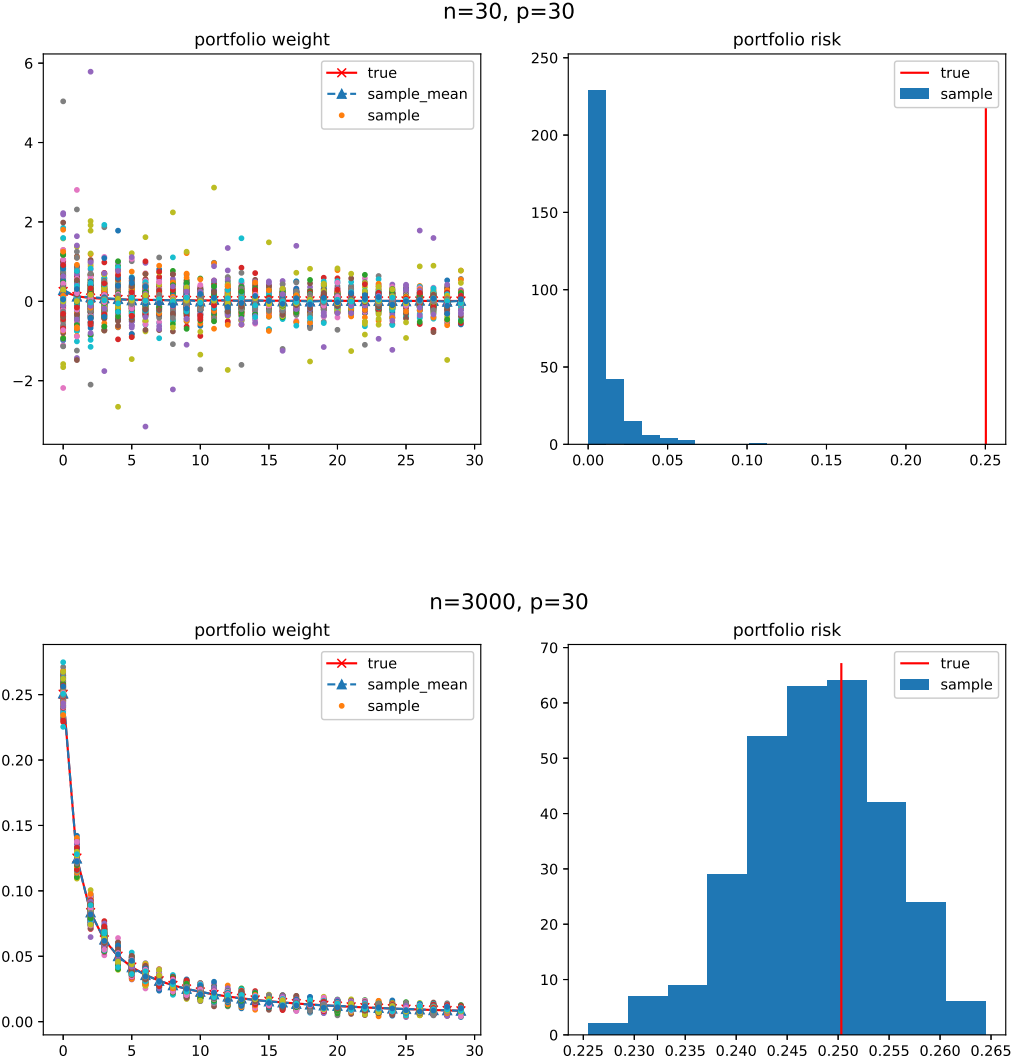}
     \end{subfigure}
     \caption{We select eigenvalues of $\Sigma$ equally spaced between 1 to 30. Namely $\sigma_k^2=k, 1\le k\le 30$. We generate 300 samples for each of the two settings $(n,p)= (30,30)$ and $(3000,30)$. when $\frac{p}{n}=1$, the error of the portfolio weight is  $O(1)$. when $\frac{p}{n}=1/100$, the error of the portfolio weight is  $O(1/10)$}
     \label{fig:diagonal}
\end{figure}

 On the left figure \ref{fig:diagonal}, true optimal weight is red  line which is closely aligned with the mean value of sample optimal weights which is show as blue connected dash-line. As we see the standard deviation in sample portfolio weight is at $O(\sqrt{p/n})$. As $p/n$ decreases, the sample portfolio weight become less volatile around the true portfolio weight. On the right, the sample optimal risk has higher chance of underestimate the true optimal risk. As $p/n$ decreases, the sample portfolio risk become less volatile and more centered around the true portfolio risk. 
%


\subsection{Second example: sample portfolio of dependent assets}

For general assets with dependence, \ref{eqn:true weight formula} and \ref{eqn:sample weight formula} have provided the formulas. Again we will only use experiments to show relations between the sample portfolio weight $\hat{w}^*$ and the true portfolio weight $w^*$.

\begin{figure}[H]
     \centering
     \begin{subfigure}[b]{0.9\textwidth}
         \centering
         \includegraphics[width=\linewidth]{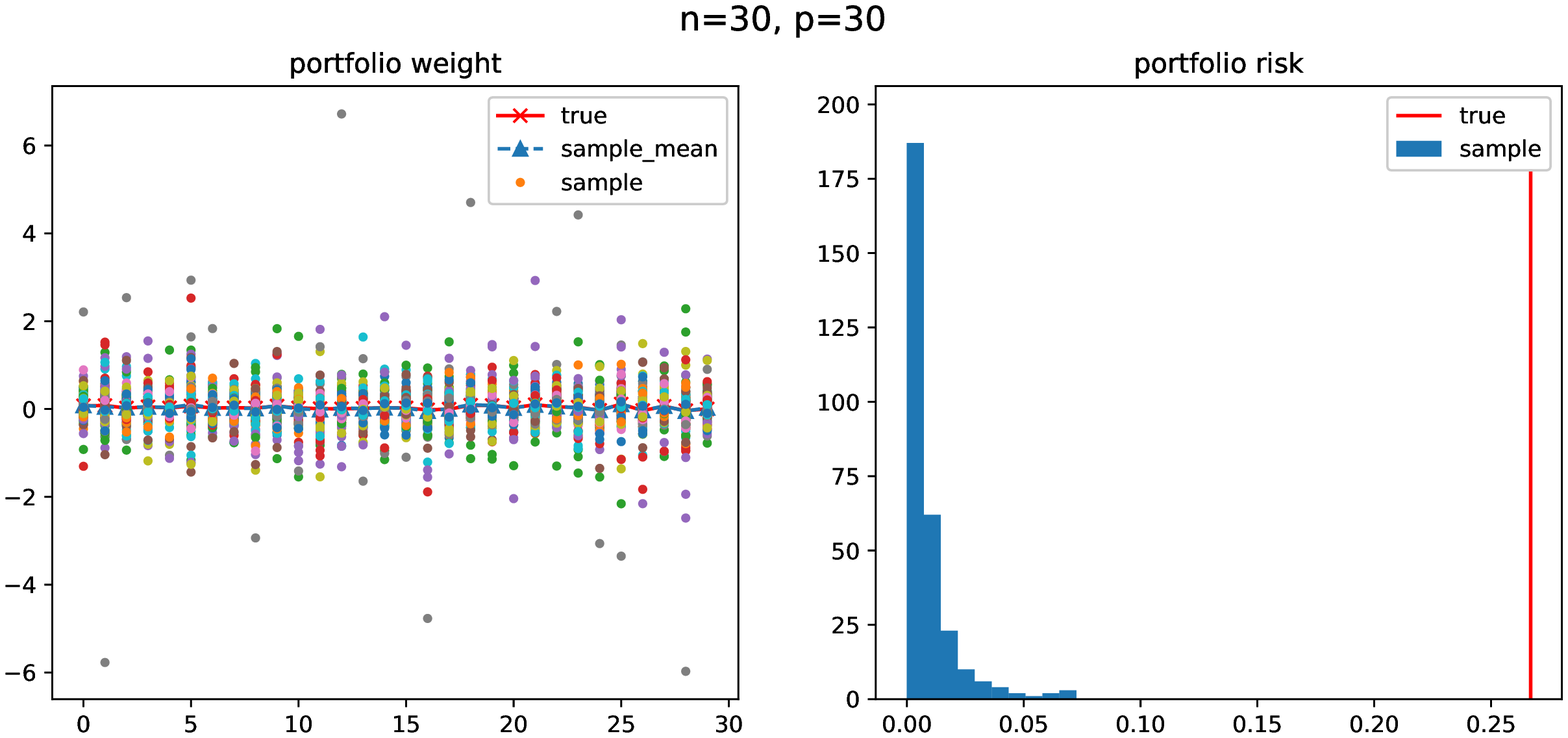}
     \end{subfigure} 
    \end{figure}%
\begin{figure}[ht]\ContinuedFloat
 \centering
     \begin{subfigure}[b]{0.9\textwidth}
         \centering
    \includegraphics[width=1\linewidth]{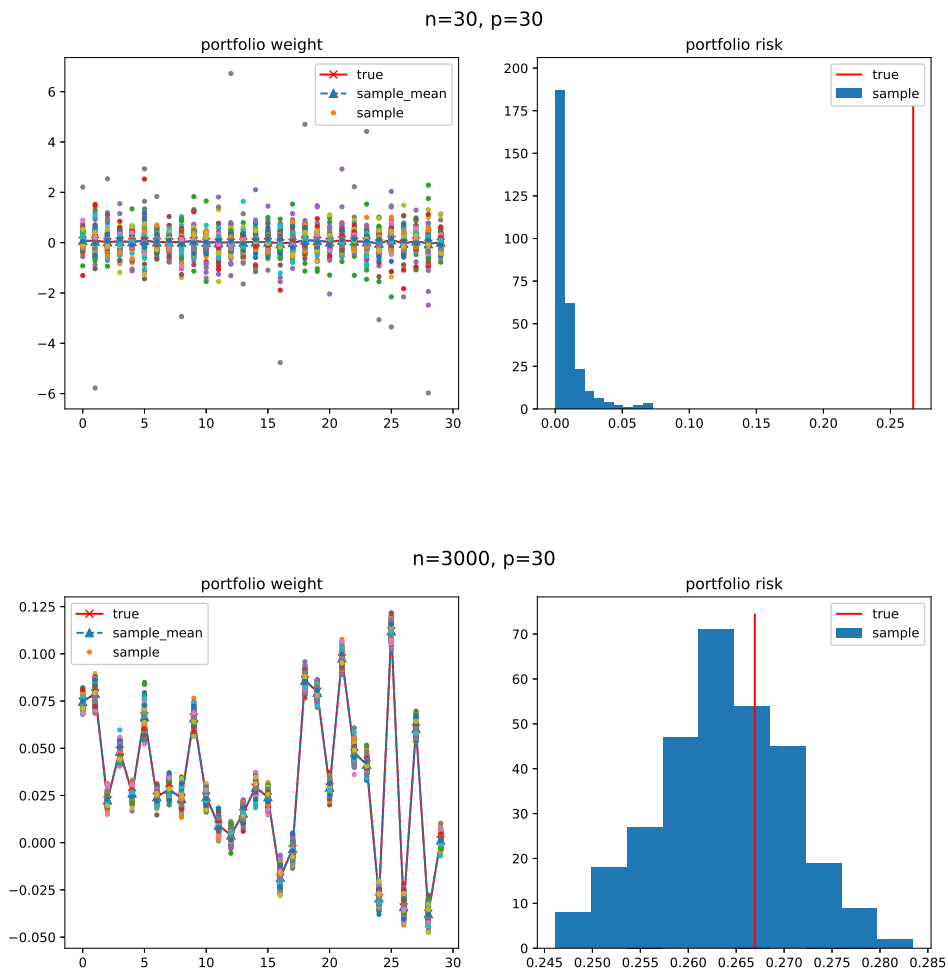}
     \end{subfigure}
     \caption{We still select eigenvalues of $\Sigma$ equally spaced between 1 to 30. Namely $\sigma_k^2=k, 1\le k\le 30$. We now select $P$ to be a random orthogonal matrix according to the Haar measure.  }
     \label{fig:dense}
\end{figure}

Since we are using non-identity orthogonal matrix $P$ to create dependence among the assets, the true optimal portfolio weight is not ordered. The concentration and deviation properties of the sample portfolio weight has not changed. On the left figure \ref{fig:dense}, true optimal weight is the red  line which is still closely aligned with the mean value of sample optimal weights which is shown as blue connected dash-line. As we see the standard deviation in sample portfolio weight is at $O(p/n)$.  On the right, the sample optimal risk has higher chance of underestimate the true optimal risk. As $p/n$ decreases, both sample weight and sample risk become more accurate.

%

\subsection{The order of error in sample optimal portfolio}
We summarize our findings from previous examples and experiments as the following conjecture 
\begin{conjecture}
Error estimates for $\hat{w}^*$ (\ref{eqn:sample weight formula}) compared with $w^*$( \ref{eqn:true weight formula}): If assume eigenvalues of true covariance matrix $\Sigma$ are $\sigma_k^2$, then
\begin{equation*}
   \E| \hat{w}^*_k - w^*_k| = O\left( \sigma_k \times \sqrt{\frac{p}{n}} \right),\quad  \forall 1\le k \le n
\end{equation*}
The constant in the order depends on smallest and largest eigenvalues of $\Sigma$.
\end{conjecture}

Even though we can not prove this in  full generality, we can show 
\begin{theorem}
Assume the true covariance of assets has diagonalization $\Sigma=P^TD^2P$ with $D= diag (\sigma_1,\cdots \sigma_p)$ and asset return data $X= \cN DP$ where $\cN$ is a $n\times p$ matrix with i.i.d. standard  Gaussian random variables (mean zero and variance one).  And the sample covariance matrix 
\[
\hat{\Sigma} = P^T D\frac{\cN^T\cN}{n} D P
\]
Then error in sample free optimal weight $S_{\hat{\Sigma}}$ of \ref{eqn:sample weight formula} satisfies the bound
\[
\E \|S_{\hat{\Sigma}} - S_{{\Sigma}}\|_2 \le  O\left( p \; \sigma_{max}\; \sigma_{min}^{-1}  \sqrt{\frac{p}{n}} \right)
\]
with high probability. where $\| \cdot \|$ is the matrix 2-norm.
\end{theorem}
\begin{proof}
From \ref{eqn:true weight formula} and \ref{eqn:sample weight formula}, we know the free optimal weights $S_{\Sigma}$ and $S_{\hat{\Sigma}}$  solves the linear system 
\begin{align*}
     \Sigma S_{\Sigma} & = P^T D^{2}P S_{\Sigma} =  \1  \\
     {\hat{\Sigma}} S_{\hat{\Sigma}} & = P^T D \left( \frac{\cN^T\cN}{n}\right) DP S_{\hat{\Sigma}}   =  \1 
\end{align*}
To compare $S_{\Sigma}$ and $S_{\hat{\Sigma}}$, we use perturbation theory of linear systems. 
Given linear system $Ax=b$ and its perturbed version $(A+B) \hat{x} =b$. then \begin{align*}
    (A+B)(\hat{x}-x +x) & = b \\
    (A+B)(\hat{x}-x) & = Ax- (A+B)x \\
    \hat{x}- x &= -(A+B)^{-1}Bx
\end{align*}

Therefore for any norm $\|\cdot \|$.
\[
\|\hat{x} -x\| \le \|(A+B)^{-1} B \| \|x \|
\]
Replace $A= P^TD^2P$ and $A+B=P^T D \left( \frac{\cN^T\cN}{n}\right) DP$, we find 
\begin{align*}
    \|S_{\Sigma} - S_{\hat{\Sigma}} \|_2 & \le \left\| P^T D^{-1} \left( \frac{\cN^T\cN}{n}\right)^{-1} D^{-1} P  P^T D \left( \frac{\cN^T\cN}{n}-I\right) DP \right\|_2 \|S_{\Sigma}\|_2 \\
    & \le \left\| P^T D^{-1} \left( \frac{\cN^T\cN}{n}\right)^{-1} \left( \frac{\cN^T\cN}{n}-I\right) DP \right\|_2 \|S_{\Sigma}\|_2 \\
\end{align*}
Notice $D$ is diagonal and $P$ is orthogonal, we see 
\[
  \|D\|_2= \sigma_{max},  \; \|D^{-1}\|_2= \sigma_{min}^{-1}, \; \|P\|_2 =1
\]
Denote $M:= \frac{\cN^T\cN}{n}$. Therefore we have the bound 
\[
\|S_{\Sigma} - S_{\hat{\Sigma}} \|_2 \le \|S_{\Sigma}\|_2  \sigma_{min}^{-1} \sigma_{max} \left\| I-M^{-1} \right\|_2 
\]
Notice 
\begin{align*}
    \|I-M^{-1}\|_2 & = \max (|1-\lambda_{min}^{-1}|, |1-\lambda_{max}^{-1}|) \\
    & \le |1-\lambda_{min}^{-1}|+ |1-\lambda_{max}^{-1}| \\
    & = \lambda_{min}^{-1} - \lambda_{max}^{-1}
\end{align*}
From random matrix theory, eigenvalues of $M$ follows Marchenko-Pastur distribution. Moreover, smallest and largest eigenvalues of $M$ satisfies (see \cite{rudelson2010non})
\[
\E \lambda_{max}(M) \le \left(1+\sqrt{\frac{p}{n}}\right)^2, \; \E \lambda_{min}(M) \ge \left(1-\sqrt{\frac{p}{n}}\right)^2
\]
It is known the non-asymptotic behavior of $\lambda_{max}$ and $\lambda_{min}$ satisfies sub-exponential tails
\[
\p \left( \left(1-\sqrt{\frac{p}{n}}\right)^2 -t\le \lambda_{min}(M) \le \lambda_{max}(M) \le \left(1+\sqrt{\frac{p}{n}}\right)^2 +t\right) \le 2e^{-\sqrt{n}t}
\]

The sub-exponential tail properties implies with high probability ($1-O(n^{-c})$) so that $\lambda_{min}\ge 1-O(\sqrt{\frac{p}{n}})$ and $\lambda_{max}\le 1+O(\sqrt{\frac{p}{n}})$ is concentrated  around $(\E \lambda_{min})^{-1}- (\E \lambda_{max})^{-1} $. Then  with high probability
\begin{align*}
    \E \|I-M^{-1}\|_2 
    &\le \E \lambda_{min}^{-1} - \E \lambda_{max}^{-1} \\
    & \le C\left[\frac{1}{1-O(\sqrt{\frac{p}{n}})} - \frac{1}{1+O(\sqrt{\frac{p}{n}})} \right] \\
    &= O(\sqrt{\frac{p}{n}})
\end{align*}
Therefore we conclude
\[
\|S_{\Sigma} - S_{\hat{\Sigma}} \|_2  \le   \|S_{\Sigma}\|_2 \; \sigma_{min}^{-1} \sigma_{max}  O(\sqrt{\frac{p}{n}})
\]
\end{proof}

Notice this result is closely related to how $\hat{w}^*$ behave.  For instance, if we assume $D=P=I$, then $\|S_{\Sigma}\|_2 = p$, we see 
\[
\|w^* - \hat{w}^* \|_2 \le \|\hat{w}^* - S_{\hat{\Sigma}}/p \|_2 +    \sigma_{min}^{-1} \sigma_{max}  O(\sqrt{\frac{p}{n}})
\]

\section{LoCoV: low dimension covariance voting}
So far we have seen that large errors are present when we use \ref{eqn:sample portfolio optimization} to approximate \ref{eqn:true portfolio optimization} especially when $p/n$ is not small. The natural question is whether there is a rescue to reduce the errors when $p$ and $n$ are comparable. The answer is positive and we provide LoCoV algorithm, low dimension covariance voting, which consistently outperform the sample optimal portfolio $\hat{w}^*$. 

Let us start with the motivation behind LoCoV. From random matrix theory, the sample covariance approaches to the true covariance as $p/n \to 0$. Suppose we have $n=30$ samples for $p=30$ assets. Then for any two assets, $X_k$ and $X_t$, the $2 \times 2$ sample covariance matrix $\hat{\Sigma}_{ks}$ for assets $X_k$ and $X_t$  has 30 samples thus feature-to-sample ratio is $2/30$ which is much smaller compared with $30/30$ for the sample covariance matrix $\hat{\Sigma}$ for all 30 assets.

On the other hand, philosophically portfolio optimization is to compare different assets and find proper investment hedges (ratios). Since we have a very accurate sample covariance matrix $\hat{\Sigma}_{kt}$ for asset $X_k$ and $X_t$, we can find accurate investment relative-weights ($u_k, u_t$), invest $u_k$ on asset $X_k$ and $u_t$ on asset $X_t$, by solving \ref{eqn:sample portfolio optimization}. As we repeat this process for any pair of two assets, we can use these low dimension covariance matrices $\hat{\Sigma}_{kt}$ to accurately construct ratios $(u_k, u_t)$ and then we utilize all $p^2$ pairs of ratios to vote on each assets and obtain a final portfolio weight vector.

\begin{algorithm}[H]
\DontPrintSemicolon
\caption{\textbf{`LoCoV-$2$'}   }\label{alg:locov-2}

 \KwData{centered asset return $X\in \R^{n\times p}$, $n, p > 0$}
 Compute sample covariance matrix
 $
  \hat{\Sigma} \gets \frac{1}{n}X^T X
 $
 
 \textbf{Initialization:} $U \gets \frac{1}{2} I$,  $V \gets 0$.\\
 \tcp*{$U$ is $p\times p$ relative-weight matrix, $V$ is $p\times 1$ free-weight vector}

\For{$i\gets1$ \KwTo $p$}{
    \tcc{\emph{1. For asset $i$ find relative-weights}}
    \For{$j\gets i+1$ \KwTo $p$}{
        Extract $2\times 2$ sub-matrix $\hat{\Sigma}_{i,j}$, and solve the 2-assets portfolio optimization
        \begin{align*}
            \begin{split}
    		&\min_u \; u^T \hat{\Sigma}_{i,j}\; u \\
    		& s.t. \;  u^T \mathbbm{1} =1
            \end{split} 
        \end{align*}
    or use formula $u=(u_1,u_2) = \hat{\Sigma}_{i,j}^{-1} \1$.
    
    $U_{i,j}\gets u_1$ \tcp*{invest $u_1$ in asset $i$ }
    
    $U_{j,i} \gets u_2$ \tcp*{invest $u_2$ in asset $j$ }
    }
     
    \tcc{\emph{2. Voting}}
    Compute free-weight by uniform voting 
    \[
    V_i \gets \frac{1}{p} \sum_{j=1}^p U_{i,j}
    \]
}

Normalize $V$  
    \[
    w \gets \frac{V}{\|V\|_{s}} = \frac{V}{\sum_{i=1}^p V_i}
    \]
    
\KwOut{$w$}
\end{algorithm}

And we can easily generalize this algorithm to that using $k\times k$  dimensional covariance and solve corresponding \ref{eqn:sample portfolio optimization} for $k$ assets instead of using $2\times 2$ low dimensional covariance. Therefore we propose the following `LoCoV-$k$' algorithm.

\begin{algorithm}[H]
\DontPrintSemicolon
\caption{\textbf{`LoCoV-$k$'} ($k\ge 3$)  }\label{alg:locov-k}

 \KwData{centered asset return $X\in \R^{n\times p}$, $n, p > 0$}
 Compute sample covariance matrix
 $
  \hat{\Sigma} \gets \frac{1}{n}X^T X
 $
 
 \textbf{Initialization:} $U \gets \frac{1}{k} \1 \1^T$,  $V \gets 0$.\\
 \tcp*{$\1$ is $p\times 1$ vector of all ones, $V$ is $p\times 1$ free-weight vector}

\For{$i\gets1$ \KwTo $p$}{
    \tcc{\emph{1. For asset $i$ find relative-weights}}
    \For{$j\gets 1$ \KwTo $p$}{
        Generate index set $I=\{i,l_1,\cdots,l_{k-1}\}$ where $l_1,\cdots l_{k-1}$ random uniformly in $ \{1,\cdots,p\}\setminus\{i\}$.
        
        Extract $k\times k$ sub-matrix $\hat{\Sigma}_{I}$, and solve the k-assets portfolio optimization
        \begin{align*}
            \begin{split}
    		&\min_u \; u^T \hat{\Sigma}_{I}\; u \\
    		& s.t. \;  u^T \mathbbm{1} =1
            \end{split} 
        \end{align*}
    or use formula $u=(u_0, u_1,\cdots, u_{k-1}) = \hat{\Sigma}_{I}^{-1} \1$.
    
    $U_{i,j}\gets \frac{1}{2}u_0+\frac{1}{2}U_{i,j}$  \tcp*{invest $u_0$ in asset $i$ }
    
    $U_{l_t,i} \gets \frac{1}{2}u_{t}+ \frac{1}{2}U_{l_t,i}, \quad \forall 1\le t \le k-1 $ \tcp*{invest $u_{t}$ in asset $l_t$ }
    }
     
    \tcc{\emph{2. Voting}}
    Compute free-weight by uniform voting 
    \[
    V_i \gets \frac{1}{p} \sum_{j=1}^p U_{i,j}
    \]
     
}

Normalize $V$  
    \[
    w \gets \frac{V}{\|V\|_{s}} = \frac{V}{\sum_{i=1}^p V_i}
    \]
    
\KwOut{$w$}
\end{algorithm}
\medskip
In LoCoV-$k$, there are several tweaks from LoCoV-$2$ in order to adapt to $k$-assets.
Every time we solve a $k$-assets portfolio optimization problem, we obtain $k$ relative weights. In order to use all $k$ weights, we initialize the relative-weight matrix $U$ with all entries being $\frac{1}{k}$. If there is a new weight generated from the computation, we take average of the existing weight and the new weight. This update will  diminish old weights which is only for convenience reading and understanding the algorithm. One could take a more delicate update on entries of $U$, for example  keep track of the total number of weights generated for each entry, and then update with an average of all weights.


\section{Simulations}
We run three experiments and select $\Sigma= I, D^2, P^TD^2P$. For each experiment, we generate 300 samples $\hat{\Sigma}$ and compute  corresponding $\hat{w}^*$ and LoCoV estimator. We plot $\hat{w}^*$ in green and LoCoV-weight in black. The experiments show LoCoV consistently outperforms the sample optimal portfolio. 
\begin{figure}[H]
     \centering
     \begin{subfigure}[b]{0.9\textwidth}
         \centering
         \includegraphics[width=\linewidth]{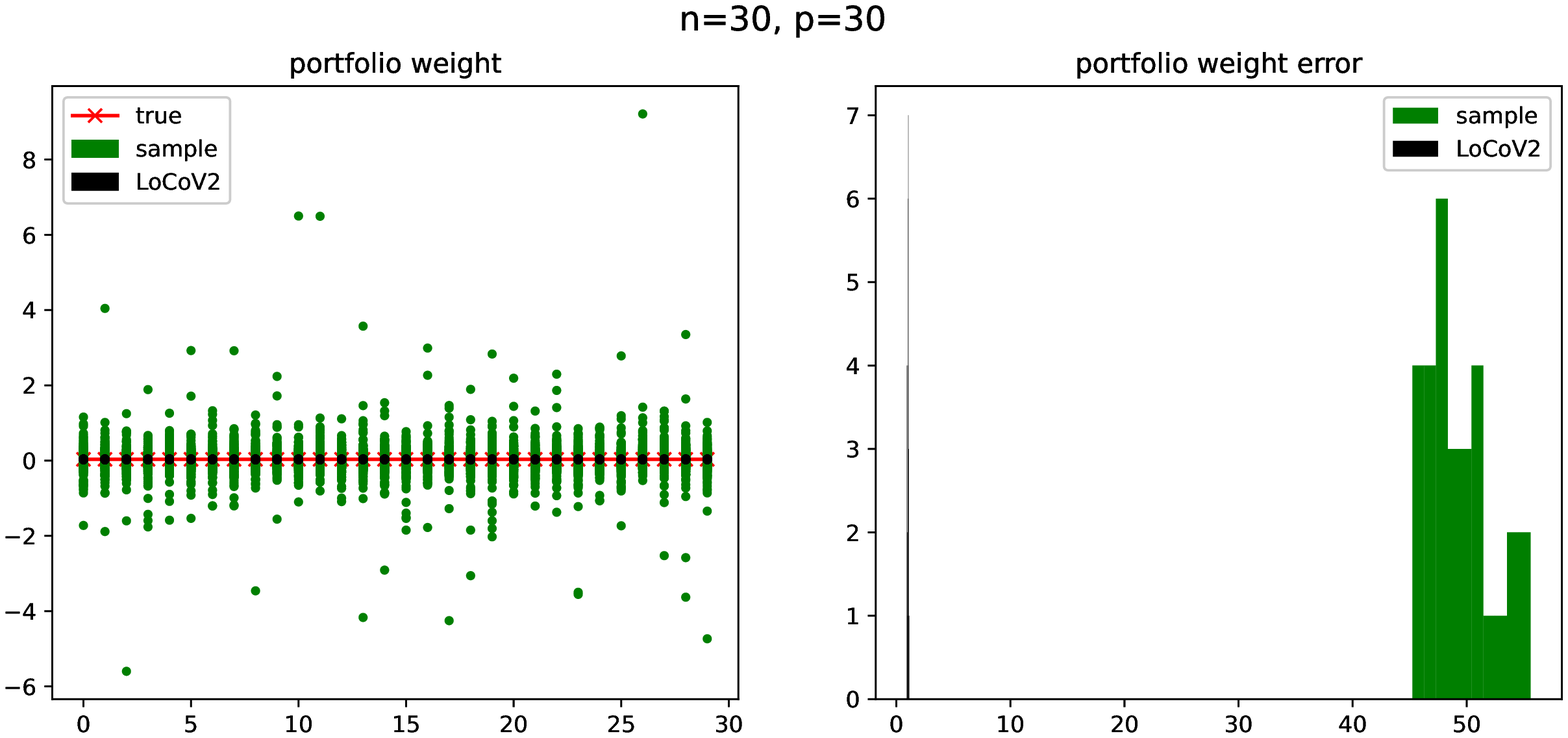}
     \end{subfigure} 
     \caption{$\Sigma=I$}
    \end{figure}%
\begin{figure}[H]
    \centering
     \begin{subfigure}[b]{0.9\textwidth}
         \centering
    \includegraphics[width=1\linewidth]{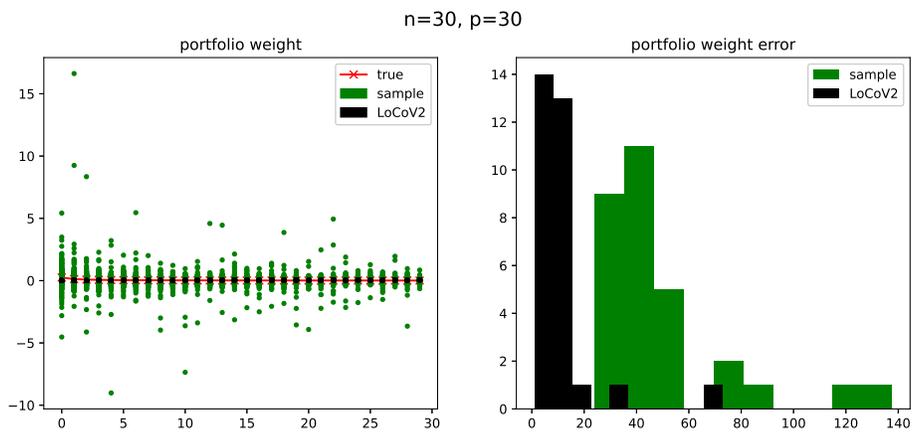}
     \end{subfigure}
     \caption{$\Sigma=D^2$ with  eigenvalues of $\Sigma$ equally spaced between 1 to 30. Namely $\sigma_k^2=k, 1\le k\le 30$.}
\end{figure}
\begin{figure}[H]
    \centering
     \begin{subfigure}[b]{0.9\textwidth}
         \centering
    \includegraphics[width=1\linewidth]{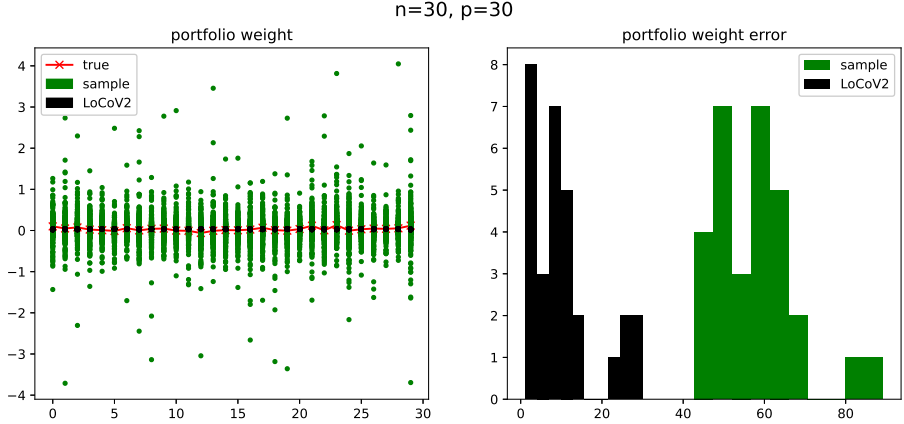}
     \end{subfigure}
     \caption{$\Sigma=P^TD^2P$ with  eigenvalues of $\Sigma$ equally spaced between 1 to 30. Namely $\sigma_k^2=k, 1\le k\le 30$. $P$ is a random orthogonal matrix according to the Haar measure.  }
     \label{fig:simulation}
\end{figure}

\section{Conclusion and open question}
We analyzed the minimum variance portfolio question with the consideration of randomness of sample covariance matrix. In light of random matrix theory, we use experiments showed the error in sample optimal portfolio has the order of the  assets-to-sample ratio $p/n$. When number of assets $p$ is not considerably smaller than the number of samples $n$, the sample optimal portfolio fails to provide accurate estimation of true optimal portfolio. Thus we proposed the LoCoV method which exploits the fact that $k$-dimensional sub-covariance matrix is more accurate thus can be used to produce relative weights among $k$ assets. Using relative weights to uniformly vote on given assets eventually improve dramatically on the performance of the portfolio.
\subsection{Adapt LoCoV to general mean-variance portfolio}
We have not discussed the role of mean return and assumed our data is centered. To adapt to general non-centered mean-variance portfolio optimization, one must modify the $k$-assets optimization sub-problem. Namely, one has to compute sample mean $\mu = \frac{1}{n} X_{i\cdot}$, and then solve the k-assets portfolio optimization
\begin{align*}
    \begin{split}
	\min_u \; & u^T \hat{\Sigma}_{I}\; u \\
	 s.t. \;  & u^T \mathbbm{1} =1 \\
	& \mu u\ge r_0
    \end{split} 
\end{align*}
where $r_0$ is the lower bound of expected return.

However, there is no guarantee to achieve the mean return $\mu w \ge r_0$ for the voting procedure produced weight $w$. Of course one can try to apply LoCoV first and check whether mean return is above the threshold $r_0$, if not then repeating the process of updating relative-weight matrix $U$ will probably improve.

\medskip

\bibliography{Portfolio}
\bibliographystyle{plain}
\end{document}